# Using Professional Social Networking as an Innovative Method for Data Extraction The ICT Alumni Index Case Study

Rasha Y. Tantawy, Ziad Farouk, Shereen Mohamed, A. H. Yousef

***Abstract*— The lack of data regarding Information and Communications Technology (ICT) sector alumni data is a known problem in several countries including Egypt. It is not clear what entry and senior jobs are occupied by alumni and which countries are attract them. This affects the planning, design and execution of both the ICT sector and the Education sector. In this research, a joint team is formulated from the Technology Innovation and Entrepreneurship Center (TIEC) and the Ministry of Higher Education. This team is undertaking extensive analysis of the structure, distribution and development of the ICT skills and employment.**

**Social Networking sites (SNS) have proved to have impact on users' lives. Famous SNS like Facebook, Twitter and LinkedIn affects the communication between humans. They are effective tools in gathering information and knowledge. This study aims to show the impact of SNS in discovering the relation between the Egyptian Universities' Graduates and the Information and Communication Technology (ICT) job Market inside and outside Egypt.**

**This research used professional social networks as a source of a significant sample of the required data. The assumption is verified with the statistics collected from the supreme Council of Universities and the Central Agency of Public Mobilization And Statistics (CAPMAS).**

**The results showed that social network can be used as a reliable information source for alumni data. It shows also that reliability increases dramatically with city of residence.**

***Index Terms*— Employment, Career, ICT, Social Networking, Knowledge Discovery**

## I. Introduction

The ICT industry is constantly evolving and new technologies are continuously being introduced. The new technologies create new ICT positions, such as Social Networking Manager, IT Architect, Knowledge Manager and Web Specialist, yet existing job descriptions such as Systems Analyst, Project Manager and Programmer still exist. ICT university alumni today can pursue a large range of ICT careers. There is an evident gap between market demand for ICT skilled caliber and their supply. This research tackles this important issue. The unavailability of alumni tracking systems causes the inexistence of data that relates the alumni with the market jobs they got. So it was not possible to relate the numbers of ICT alumni from Egyptian universities and their employment track or the countries they moved to. Neither the universities nor the syndicates have a real up to date record of those alumni. Finding a solution for this problem will enable ICT students to be aware of the possible ICT career opportunities and career tracks available. They will be also informed about the job descriptions of a computing professional. This will minimize the gap between the supply of highly educated youth and the job creation process. In future versions of the study, the skills and competencies will be considered as well.

The paper is organized as follows: Section II contains a literature review about similar work. Section III presents the background of social networks. Section IV presents the methodology and section V represents the study of the relation between the Egyptian Universities Graduates and the ICT job Market inside and outside Egypt. The conclusion is presented in Section VI.

## II. Literature Review

ICT skills of different levels of complexity are widely used throughout the economy. ICT specialists (e.g. computer professionals, electronic engineers) represent a high share of employment. The importance of ICT-skilled employment at the specialist level has increased over time. ICT-skilled employment is associated with higher levels of productivity.

At sector level, a large share of ICT skilled employment is associated with a high level of value added per employee, a sign that the use of ICT-skilled workers is associated with measurable economic benefits. The distribution of ICT skills throughout the economy is important for competitiveness, as effective use of ICTs in production and business processes is of crucial importance for countries' productivity and growth.

The need for ICT skills can be satisfied in part through education and training. ICT-related degrees with basic skills can be obtained through formal education. For specialized skills, sector-specific training and certification schemes may be crucial، given the rapid changes in skills needs and the constant introduction of new technologies.

Egypt is considered a good place to provide outsourcing. Skill needs can also be satisfied also by migration and acquisition between Egyptian companies and international ones. Egypt has encouraged inflows of international IT companies and many Egyptian skills migrated permanently or temporarily to Rasha Y. Tantawy is with the Technology Innovation and Entrepreneurship Center, Egypt (e-mail: rTantawy@tiec.gov.eg).

Ziad Farouk, Sherin Mohamed and Ahmed Hassan Yousef, is with ICTP, Ministry of Higher Education, Egypt.

other countries for better compensations. Internet recruitment is a new way of satisfying changing skills needs at firm level. It appears to be relatively more important and increasing in ICT-related fields and sectors.

The challenge is to develop, nurture and attract talents as well as to strengthen human capital investment. The current main issues encompass, on the demand side, clearly defining what e-business and ICT skills are needed, thereafter enabling forecasting and scenarios exercises. On the supply side, the challenges lie in the provision of a sufficient volume of skilled labor, with accurate and up-to-date knowledge that matches the demand requirements. Last but not least, the supply of talented and skilled people needs to be scalable and sustainable over-time [1].

Employment growth is being driven by expansion in many companies, by a continuing flow of inward investment, and by the formation of new startups. There is a pattern of simultaneous creation and loss of jobs, where lower-skilled jobs are being replaced with higher-skilled jobs [2].

According to the World Bank survey (2007), only 6.6% of enterprises surveyed have a major problem with skills employment in Egypt, moving up the ranking significantly from 2004 when it was 29.8%. This indicates the importance of tracking ICT alumni and monitoring the demand and supply of skills in the sector.

There were pervious researches done by other countries, to rein the numbers of the demands and offers in the ICT job Market from employment sites like www.seek.com.au.

Another research was made to rein the number of free lancers among 156 countries. Egypt is ranked at the 14th spot with total of 2445 free workers serve 320 clients in 156 countries as mentioned in www.techservealliance.org.

From previous researches, it is insured that it is necessary to have a guide that links the ICT graduates with the job Market in order to develop the courses offered in this field and develop the students' skills to meet the job Market requirements.

## III. Social Network Sites

Internet has changed the world rapidly; it changed the professional, personal and social life of humans. communication is one of the most important benefits from the use of the internet [4]. The limitless connectivity and potential to create an open social order and system of interaction and collaboration have been made possible only because of ICT. We can see the impact of ICT in every walk of life [5].

Social network is a broad term used to denote the blogs, user created videos and wikis. A social networking is an online service, platform or site that focuses on building and reflecting of social network or social relations among people who share interests and activities. Social networking often involves grouping specific individuals or organizations together. Social network provides a quick, low technology method to generate, maintain web based subject guides. Most social network promotes free flow of information and sharing of resources beyond boundaries [5].

Social networking sites are a web provision where millions of people can join together to form an online community and hence, millions of communities form a social network to share knowledge, information and even culture. The idea of social networking originated in 1995 and gave birth to an early social network called Classmates. This was created to keep students in connection even after leaving the school or class (Classmates.com). In 1997, another SNS "SixDegree.com" was released and then this development carried on with the emergence of other social networking sites, such as: Cyworld (2001), Friendster (2002), LinkedIn (2003), Myspace (2005), Yahoo 360 (2005), Twitter (2006) and Facebook (2006) [4].

These social networking sites have converted the world from a global village to a social global village or a social globe. People can communicate with others while sitting in our room and look for their networks [3]. Users prefer social networking sites to access information as it reduces physical strain, save the time; they are able to complete the work within time, minimize expenses and keep accuracy.

Facebook for example accumulated almost 700 Million users worldwide as of June 2011. If we were to in terms of user growth on Facebook, Egypt is ranked at the 8th spot. With that information, SNS are creating huge opportunities for people to have leverage on each other [3]. LinkedIn Founded in December 2002 and launched in May 2003, it is mainly used for professional networking. As of 9 February 2012, LinkedIn reports more than 150 million registered users in more than 200 countries and territories. It is reported that LinkedIn has 21.4 million monthly unique U.S. visitors and 47.6 million globally.

This study aims to configure the importance of SNS in gathering information. The importance of the study is to know the Market map of the Egyptian Universities' Graduates in the ICT field in order to know the demand of the job Market inside and outside Egypt. It is also beneficial to know the countries that need the Egyptians Graduates and on what basis. This could be useful to put an image on how to develop the study of ICT in the Egyptian colleges and what skills should the students gain from their study to meet the needs of the Egyptian job Market and the Global job Market as well.

The source of information was based on the LinkedIn which allows the users to register their personal information including the faculty and university from where they earn their education. They also register the companies where they work, their jobs and in which country.

## IV. Methodology

This pilot study tracked ICT graduates through social networks. The study is based on extracting information from social networks that ICT professionals and graduates use to communicate with each other. Information posted in their accounts include; year of graduation, graduating university and employer, post and employment history. Out of the many social networks available, LinkedIn is used as a source of information because it is a popular professional social network amongst the ICT alumni and workers. It is considered as a credible source of information. Because there was no API supported with this site so, the data was extracted manually.

It is assumed that the Linked In professional social network is used a reasonable sample of ICT alumni to record their information in a professional manner, for professional use. We assume also that the information posted by the user is correct and credible. The following precautions must be taken into account: The data of social networks users may be incomplete. One user may have multiple accounts which can give misleading statistical numbers. Many professionals do not have accounts in social networks. Most of user accounts remain active even after death.

According to the national statistics, the total number of ICT graduates in 2008 is 7,400 graduates. The total number of students in the interval 2007-2012 is estimated to be 37,000 graduates. It is estimated that the interval between 1970 to 2011 has about 140,000 to 170,000 alumni (The number was extrapolated from the Annual Bulletin of the Central Agency for Public Mobilization and Statistics) [3].

The pilot study was conducted on eight Egyptian Universities; Cairo University, Ain Shams University, Helwan University, Alexandria University, Tanta University, Suez Canal University, Assuit University and Mansoura University. The universities were chosen to be geographically distributed around Egypt. Information regarding graduates of these universities during the interval 2007- 2011 and 1970- 2011 is extracted from Linked In. The former period depicts the newly graduates while the latter depicts graduates from the start of the formation of the ICT sector. This pilot study should give us a panoramic view of the ICT graduates in Egypt.

## V. Results And Discussions

**Table I** shows the aggregated results of the pilot study for eight Egyptian universities. As shown the total number of registered ICT graduates on LinkedIn from 1970 to 2011 is 16,561 representing 10- 12% of the total graduates.

TABLE I
TOTAL NO OF ICT GRADUATES 1970-2011

| **University** | **Total no of graduates on LinkedIn 1970-2011** | **Total no of graduates on LINKEDIN 1970-2011 (*working in Egypt*)** |
|---|---|---|
| Cairo | ٧،٠٣١ | ٤،٦٨٨ |
| Ain Shams | ٤،٠٦٢ | ٢،٧٣٥ |
| Alexandria | ٢،٣٨٠ | ١،٢٨٥ |
| Suez Canal | ٤٧٩ | ٣٤٠ |
| Tanta | ٣٠٤ | ٢١٤ |
| Mansoura | ٧٨٤ | ٤٨١ |
| Helwan | ١،٢٢٦ | ٩١٩ |
| Assuit | ٢٩٥ | ١٦٠ |
| TOTAL | 16,561 | 10,822 |

The percentage of those who work in Egypt during the interval 2007-2011 is 86%. Table II shows the recent data of alumni during the interval 2007 – 2011. The percentage of registered users equals 13.4%. The percentage of alumni working in Egypt decreases for senior positions during the interval 1970-2011 to reach 65% as shown in figure 1. This means that 35% are distributed worldwide; the United States being the primary attraction with 31%. USA is followed by the UAE, Saudi Arabia and Canada as shown in figure 2.

TABLE II
ICT GRADUATE DISTRIBUTION IN EGYPTIAN UNIVERSITIES

| **University** | Number of ICT alumni 07-11 | Number of ICT alumni 07-11 with profiles on Linked In | Number of ICT alumni 07-11 employed in Egypt |
|---|---|---|---|
| Cairo | 10,303 | 2,516 | 2,155 |
| Ain Shams | 7,438 | 1154 | 1037 |
| Alexandria | 5,835 | 668 | 544 |
| Suez Canal | 1,826 | 141 | 127 |
| Tanta | 1,406 | 97 | 83 |
| Mansoura | 4,234 | 278 | 223 |
| Helwan | 6,953 | 407 | 372 |
| Assuit | 2,078 | 105 | 70 |
| TOTAL | ٤٠،٠٧٢ | ٥،٣٦٦ | ٤،٦١١ |

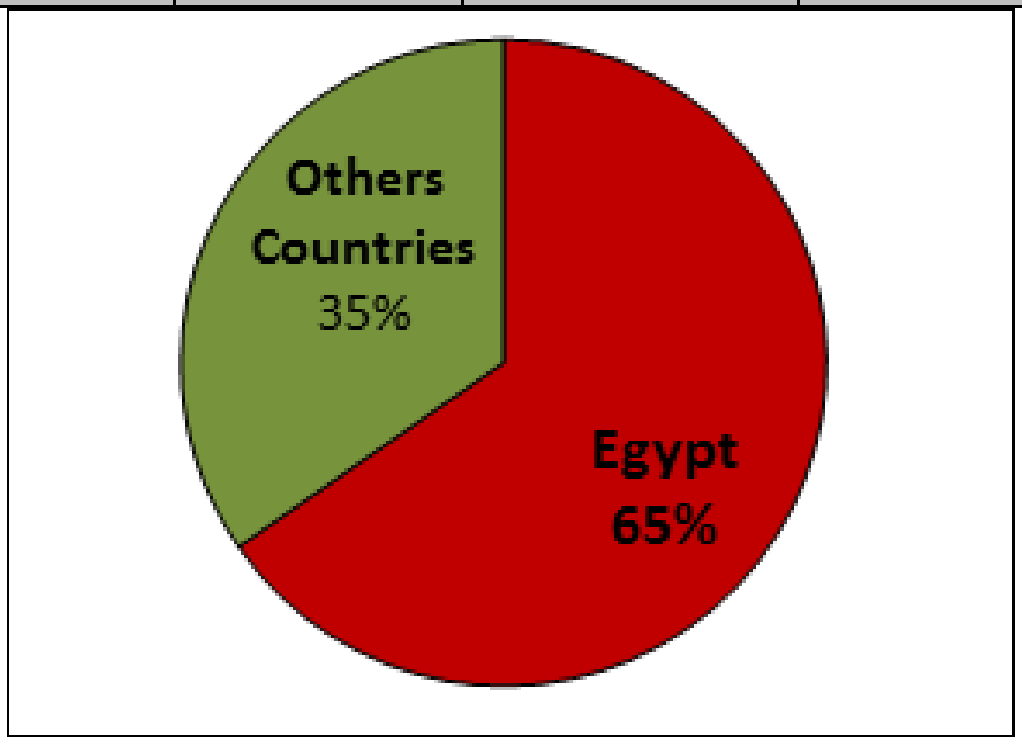

Figure 1. 1970-2011 Alumni Distribution on Countries of Residence

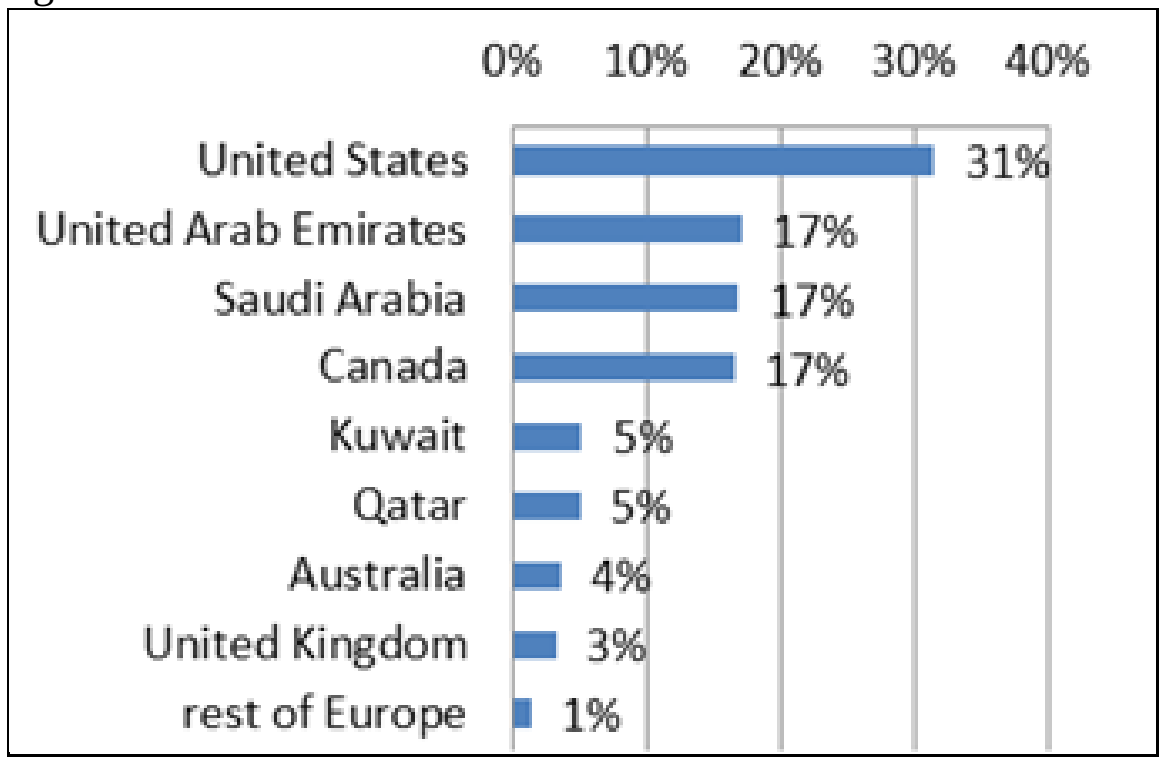

Figure 2. Distribution of Egyptian Alumni working outside Egypt

Figure 3 shows the number of Egyptian ICT graduates with LinkedIn profiles that are employed in other countries according to their graduating university. The greatest numbers are from Cairo University, followed by Ain Shams and Alexandria Universities. The biggest attractor to ICT graduates is the United States of America, followed by Canada and the United Arab Emirates consecutively. What is worth noting that Saudi Arabia and UAE are the biggest attractors for alumni from universities located outside Cairo and Alex This may be due to the fact that graduates from universities in the first and second capitals of Egypt are more language fluent than their counterparts in less central cities. It should be noted that the main attractors are English speaking countries, which is the number one language spoken in Egypt after the mother language Arabic.

From another perspective, when taking a closer look at the ICT graduates specific to the term 2007 to 2011, it was found that the Cairo University ICT graduates are the highest participators on LinkedIn (24%) followed by graduates of Ain

Shams University (16%) and then Alexandria University (11%) as shown in **figure (5)**. The results as conveyed with what is previously discussed.

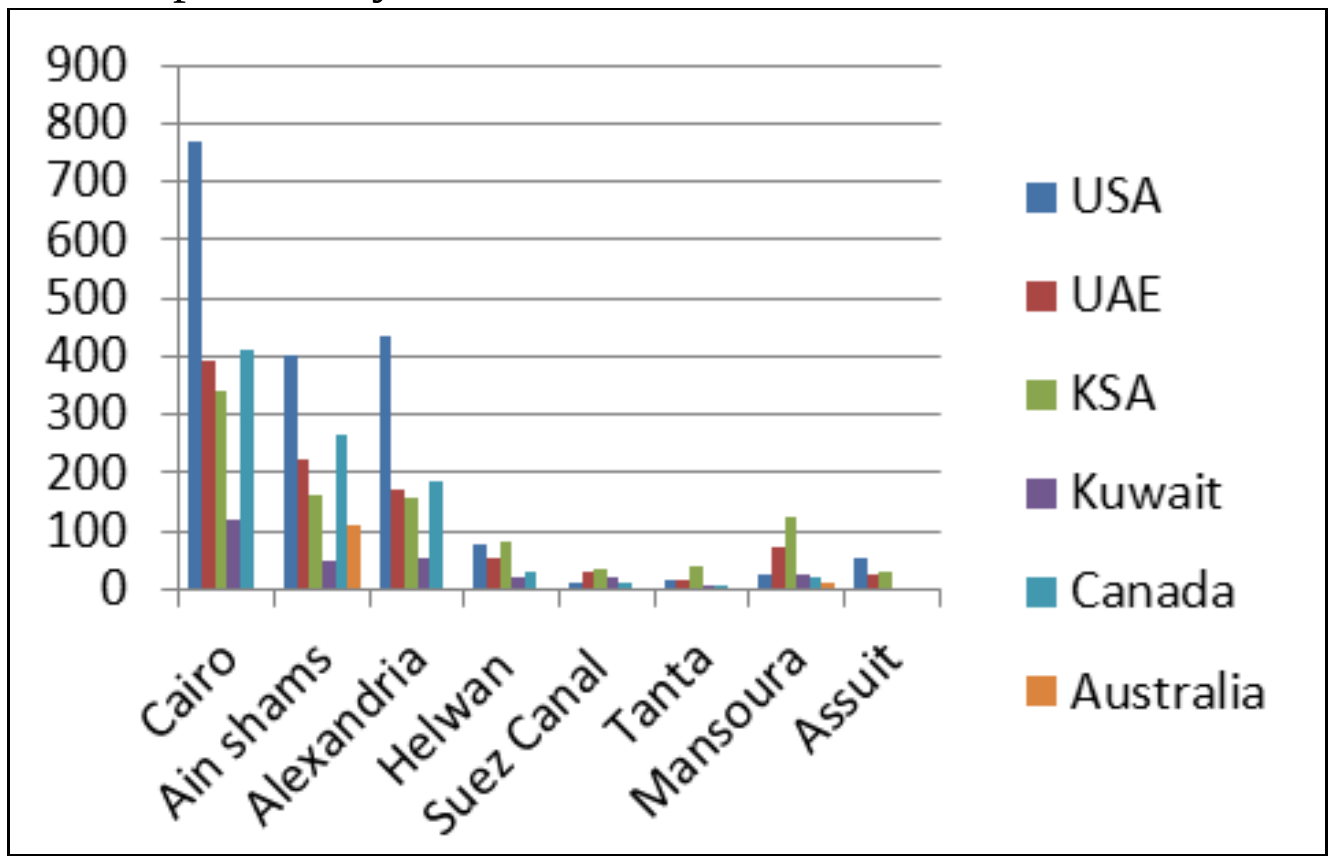

Figure 3. Countries that employ Egyptian ICT graduates (by university)

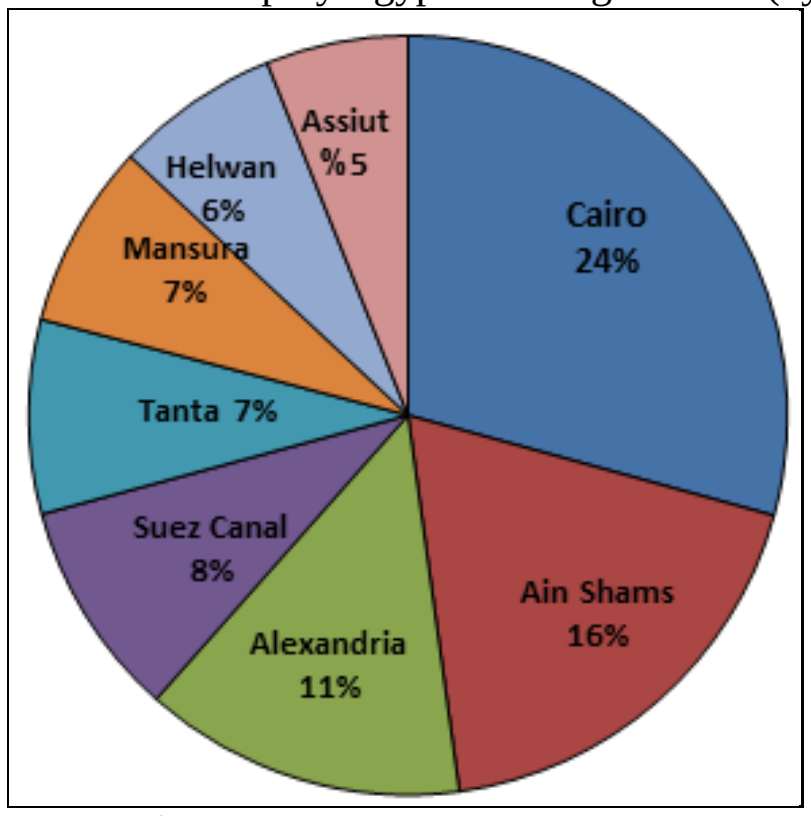

Figure 4. Percentage of ICT Graduates during 2007-2011 with LINKEDIN profiles per university

**Figure 5** shows where Egypt's ICT graduates are mostly employed. HP is the top employer followed by IBM and then ITWORX. The results also showed that ICT graduates working in governmental institutions are less keen to register on Linked In, contrary to those working in the private sector due to the need more networking capabilities.

## VI. CONCLUSION

The research has shown that professional social networks are a good source of information regarding ICT graduates from Egyptian Universities specifically those in the main cities such as Cairo and Alex. Since this source, though credible, only covers about 10- 12% of alumni of the tested period, it is therefore suggested that another methodology is to be implemented in the next stages of the study.

The research has also revealed the migration of Egyptian ICT alumni to other countries that shows the increased employability of ICT graduates.

## ACKNOWLEDGEMENT

The authors would like to acknowledge the support done by the Ministry of Communication and Information Technology and the Ministry of Higher Education.

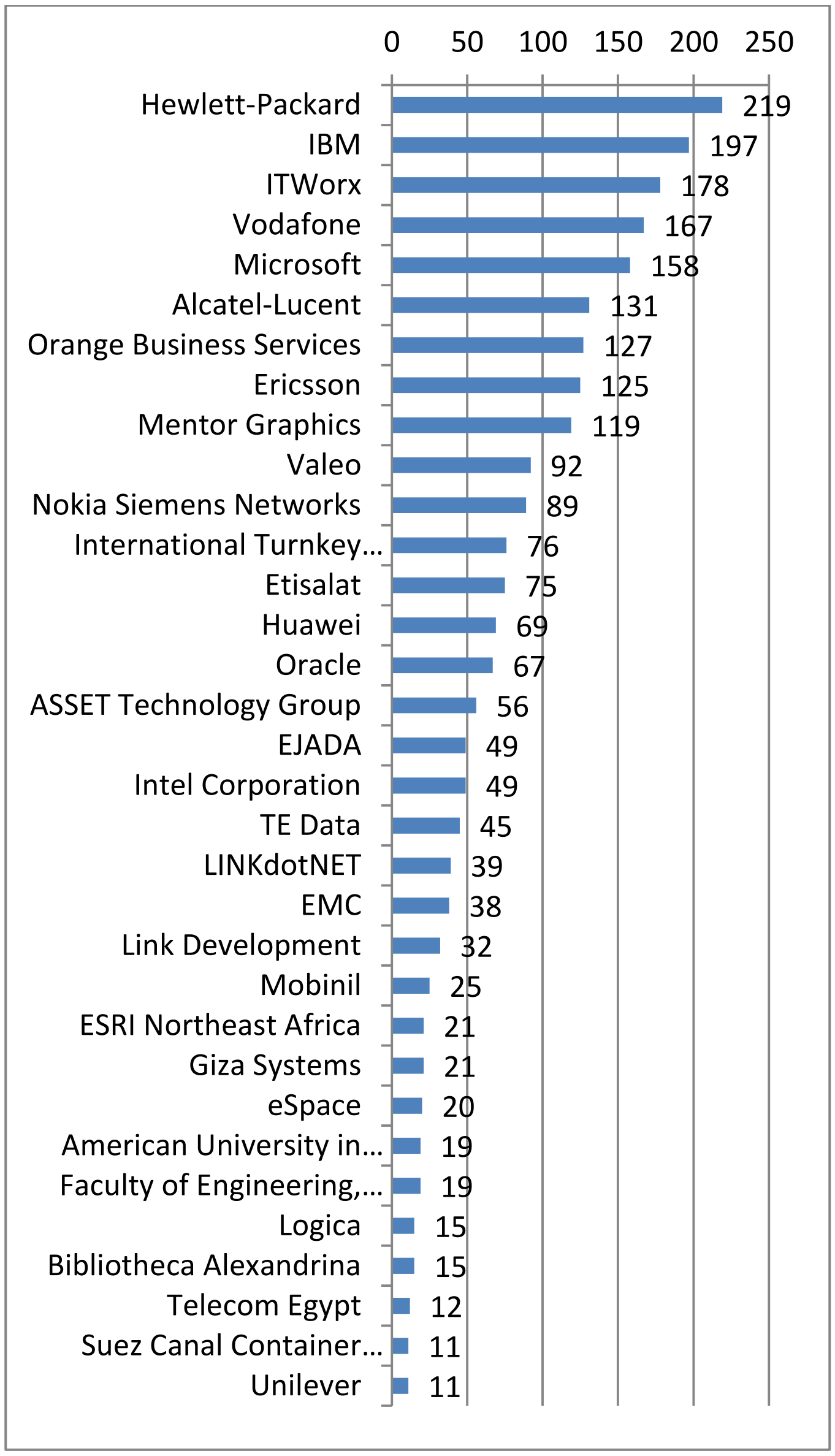

Figure 5. Distribution of Egyptian Alumni according to Employers